\documentclass[10pt,twocolumn]{article}
\usepackage{graphicx,url,array}
\usepackage[left=1.4cm,right=1.4cm,bottom=2.4cm,top=2.1cm]{geometry}
\usepackage{flushend,caption}
\usepackage{cite}
\usepackage{amsmath,amssymb,amsfonts,bm}
\usepackage{algorithmic}
\usepackage{graphicx}
\usepackage{textcomp}
\usepackage{xcolor}
\usepackage{url}
\usepackage{multirow}
\usepackage{subfig}
\let\OLDthebibliography\thebibliography
\renewcommand\thebibliography[1]{
  \OLDthebibliography{#1}
  \setlength{\parskip}{0pt}
  \setlength{\itemsep}{6pt plus 0.3ex}
}
\def\BibTeX{{\rm B\kern-.05em{\sc i\kern-.025em b}\kern-.08em
    T\kern-.1667em\lower.7ex\hbox{E}\kern-.125emX}}
\begin{document}
\title{Sound Event Detection with Depthwise Separable and Dilated Convolutions}
\author{Konstantinos Drossos$^{\dagger}$, Stylianos I. Mimilakis$^{\ddagger}$, Shayan Gharib$^{\dagger}$,\\ Yanxiong Li$^{\star}$, and Tuomas Virtanen$^{\dagger}$\vspace{18pt}\\
$^{\dagger}$Audio Research Group, Tampere University, Tampere, Finland\\
$^{\ddagger}$Semantic Music Technologies Group, Fraunhofer-IDMT, Ilmenau, Germany\\
$^{\star}$School of Electronic \& Information Engineering, South China University of Technology,\\
Guangzhou, China
}
\date{}
\twocolumn[
\maketitle
  \begin{@twocolumnfalse}
    \maketitle
        \begin{abstract}
        State-of-the-art sound event detection (SED) methods usually employ a series of convolutional neural networks (CNNs) to extract useful features from the input audio signal, and then recurrent neural networks (RNNs) to model longer temporal context in the extracted features. The number of the channels of the CNNs and size of the weight matrices of the RNNs have a direct effect on the total amount of parameters of the SED method, which is to a couple of millions. Additionally, the usually long sequences that are used as an input to an SED method along with the employment of an RNN, introduce implications like increased training time, difficulty at gradient flow, and impeding the parallelization of the SED method. To tackle all these problems, we propose the replacement of the CNNs with depthwise separable convolutions and the replacement of the RNNs with dilated convolutions. We compare the proposed method to a baseline convolutional neural network on a SED task, and achieve a reduction of the amount of parameters by 85\% and average training time per epoch by 78\%, and an increase the average frame-wise F\textsubscript{1} score and reduction of the average error rate by 4.6\% and 3.8\%, respectively. 
        \vspace{6pt}
        \end{abstract}
        \vspace{18pt}
  \end{@twocolumnfalse}
]
\section{Introduction}
Sound event detection (SED) is the task of identifying onsets and offsets of target class activities in general audio signals~\cite{cakir:2017:taslp}. A typical SED scenario involves a method which takes as an input an audio signal, and outputs temporal activity for target classes like ``car passing by'', ``footsteps'', ``people talking'', ``gunshot'', etc~\cite{cakir:2017:taslp,drossos:2019:dcase}. The time resolution of the activity of classes can vary among different methods and datasets, but typically is used 0.02 sec~\cite{cakir:2017:taslp,drossos:2019:dcase,kapka:2019:dcase,grondin:2019:dcase}. Also, activities of classes can overlap (polyphonic SED) or not (monophonic SED). SED can be employed in a wide range of applications, like wildlife monitoring and bird activity detection~\cite{cakir:2017:eusipco,adavanne:2017:eusipco}, home monitoring~\cite{turpault:2019:dcase,pages:2017:homesound}, autonomous vehicles~\cite{xu:2017:dcase,lee:2017:dcase}, and surveillance~\cite{marchegiani:2018:listening,fu:2019:ijcai}. 

Current deep learning-based SED methods can be viewed as a composition of three functions. The first function is a feature extractor, usually implemented by convolutional neural network (CNN) blocks (i.e. a CNN followed by a non-linearity, and normalization and sub-sampling processes), which provides frequency shift invariant features of the input audio signal~\cite{cakir:2017:taslp}. The second function, implemented by a recurrent neural network (RNN), models long temporal context and inter- and intra-class patterns in the output of the feature extractor (i.e. the first function)~\cite{drossos:2019:dcase}. Finally, the third function, which is an affine transform followed by a sigmoid non-linearity (in the case of polyphonic detection), performs the classification. In~\cite{cakir:2017:taslp} is described a widely adopted method that conforms to the above mentioned scheme, consisting of three CNN blocks followed by an RNN and a classifier. This method is termed as convolutional recurrent neural networks (CRNN) and has been used in a variety of audio processing tasks, like music emotion recognition~\cite{malik:2017:smc}, sound event detection and localization~\cite{adavanne:2019:jstsp}, bird activity detection~\cite{cakir:2017:eusipco,adavanne:2017:eusipco}, and SED~\cite{cakir:2017:taslp}. 

The typical amount of parameters of the CRNN is around 3.5 M, and the sequence length of the input audio and the output predictions is 1024 frames. Because an RNN is used, the CRNN method cannot be parallelized (i.e. split between different processing units, e.g. GPUs). The 1024 time-frame length of the output sequence can be considered long enough to create computational problems at the calculation of the gradient, due to the RNN (e.g. gated recurrent units, GRU, or long short-term memory, LSTM). Reduction of the number of parameters of an SED model would allow the method to be fit for systems with restricted resources (e.g. embedded systems) and the training time would decrease (resulting in faster experimentation and optimization). Also, removing the RNN would allow the method to be split between different processing units, would have more efficient training, and the amount of parameters could be further reduced.  

In this paper we propose the replacement of the CNNs and the RNN. In particular, we propose the employment of depthwise separable convolutions~\cite{sifre:2014:phd,guo:2018:bmvc,howard:2017:arxiv,chollet:2017:cvpr} instead of typical CNNs, resulting in a considerable decrease of the parameters for the learned feature extractor. Then, we also propose the replacement of the RNN with dilated convolutions~\cite{holschneider:1990:springer,shensa:1992:tsp,yu:2016:MCA}. This allows modeling long temporal context, but reduces the amount of parameters, eliminates the gradient problems due to the usually long employed sequences (e.g. 1024-frame long), and allows for parallelization of the model~\cite{lea:2016:eccv,he:2019:jop}. 

Similar approaches have been proposed in~\cite{yanxiong:2019:arxiv} and in the code of the YAMNET system, available online\footnote{\url{https://github.com/tensorflow/models/tree/master/research/audioset/yamnet}}. Specifically, in~\cite{yanxiong:2019:arxiv} is proposed a method using a series of dilated convolutions as a feature extractor, instead of CNNs. The output of the last dilated convolution is given as an input to an RNN, which does not lift any of the shortcomings of using RNNs in SED. YAMNET is based on the VGG architecture, using depthwise separable convolutions. The amount of parameters of the YAMNET amounts to 3.7M and there was not a specific module for taking into account the modeling of the longer temporal context in the input audio (e.g. like an RNN or a dilated convolution).

To evaluate the impact of our proposed changes, we employ a typical method for SED that is based on stacked CNNs and RNNs~\cite{cakir:2017:taslp}, and a freely available SED dataset, the TUTSED Synthetic 2016~\cite{tutsed}. Our results show that with our proposed changes we reduce the amount of parameters by 85\% and the average time per epoch need for training by 78\% (measured on the same GPU), while we increase the frame-wise F\textsubscript{1} score by 4.6\% and decrease the error rate by 3.8\%. The rest of the paper is as follows. In Section~\ref{sec:baseline-approach} we briefly present  the baseline approach and in Section~\ref{sec:proposed-approach} is our proposed method. Section~\ref{sec:evaluation-setup} explains the evaluation set-up of our method and the obtained results are presented in Section~\ref{sec:results}. Section~\ref{sec:conclusions} concludes the paper and proposes future research directions. 

\section{Baseline approach}\label{sec:baseline-approach}
The baseline approach accepts as an input a series of $T$ audio feature vectors $\mathbf{X}\in\mathbb{R}^{T\times N}$, with each vector having $N$ features, and associated target output corresponding to the activities of $C$ classes $\mathbf{Y}\in\{0,1\}^{T\times C}$. $\mathbf{X}$ is given as an input to a learnable feature extractor $f_{\text{cnn}}$, consisting of cascaded 2D CNN blocks. Each block has a 2D CNN followed by a non-linearity, a normalization process, and a feature sub-sampling process. The output of $f_{\text{cnn}}$ is given as an input to a temporal pattern identification module $f_{\text{rnn}}$, which consists of a GRU RNN. $f_{\text{rnn}}$ is followed by a classifier $f_{\text{cls}}$, which is an affine transform followed by a sigmoid non-linearity. The output of $f_{\text{cls}}$ for each of the $T$ feature vectors is the predicted activities for each of the $C$ classes $\hat{\mathbf{Y}}\in [0, 1]^{T\times C}$. During inference process, the activities $\hat{\mathbf{Y}}$ are further binarized using a threshold of 0.5. 

\subsection{Learnable feature extractor based on CNNs}
\begin{figure}
    \centering
    \includegraphics[width=.9\columnwidth]{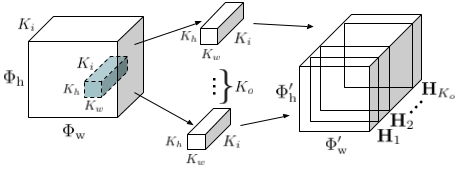}
    \caption{A typical process for a CNN. Each of the $K_{\text{o}}$ kernels, of size $K_{\text{i}}\times K_{\text{h}}\times K_{\text{w}}$, is convolved with $K_{\text{i}}$ input matrices of $T\times N$ size. The output is $K_{\text{o}}$ different matrices of $T'\times N'$ size. For clarity bias is neglected.}
    \label{fig:typical-conv}
\end{figure}

The learnable feature extractor of the baseline approach consists of three CNN blocks, each block having a typical 2D CNN followed by a rectified linear unit (ReLU), a batch normalization process, and max-pooling operation across the dimension of features. A typical 2D CNN consist of $K_{\text{o}}$ kernels $\mathbf{K}\in\mathbb{R}^{K_{\text{i}} \times K_{\text{h}} \times K_{\text{w}}}$ and bias vectors $\mathbf{b}\in\mathbb{R}^{K_{\text{i}}}$, where $K_{\text{i}}$ and $K_{\text{o}}$ are the number of input and output channels of the CNN, and $K_{\text{h}}$ and $K_{\text{w}}$ are the height and width of the kernel of each channel. Each kernel $\mathbf{K}$ is applied to the input $\mathbf{\Phi}\in\mathbb{R}^{K_{\text{i}}\times\Phi_{\text{h}}\times\Phi_{\text{w}}}$ of the 2D CNN to obtain the output $\mathbf{H}\in\mathbb{R}^{K_{o}\times\Phi_{\text{h}}'\times\Phi_{\text{w}}'}$ of the 2D CNN, as

\begin{align}\label{eq:conv}
    \mathbf{H}_{k_{\text{\text{o}}},\phi_{h}'-K_{\text{h}}, \phi_{w}'-K_{\text{w}}} =&(\mathbf{K}_{k_{\text{o}}}*\mathbf{\Phi})(k_{\text{i}},\phi_{\text{h}}-k_{\text{h}}, \phi_{\text{w}}-k_{\text{w}})\nonumber\\
    =&\sum\limits_{k_{\text{i}}}^{K_{\text{i}}}\sum\limits_{k_{\text{h}}}^{K_{\text{h}}}\sum\limits_{k_{\text{w}}}^{K_{\text{w}}}\mathbf{\Phi}_{k_{\text{i}}, \phi_{\text{h}}-k_{\text{h}}, \phi_{\text{w}}-k_{\text{w}}}\mathbf{K}_{k_{\text{\text{o}}}, k_{\text{h}}, k_{\text{w}}}\text{,}
\end{align}

\noindent
where $*$ is the convolution operator with \textit{unit stride} and zero padding. The above application of $\mathbf{K}$ onto $\mathbf{\Phi}$ leads to learning and extracting spatial and cross-channel information from the input features $\mathbf{\Phi}$~\cite{guo:2018:bmvc}, and has a computational complexity of $O(K_{\text{o}} \cdot K_{\text{i}} \cdot K_{\text{h}} \cdot K_{\text{w}}\cdot\Phi_{\text{h}}\cdot\Phi_{\text{w}})$~\cite{guo:2018:bmvc,howard:2017:arxiv,chollet:2017:cvpr}. Additionally, the amount of learnable parameters of the 2D CNN (omitting bias) is $K_{\text{i}} \cdot K_{\text{o}} \cdot K_{\text{h}} \cdot K_{\text{w}}$. Figure~\ref{fig:typical-conv} illustrates the above operation. 

In each CNN block of the feature extractor, the output of the 2D CNN is followed by ReLU, batch normalization, and max-pooling operations. The output of the max-pooling operation is given as an input to the next CNN block. The output of the third CNN block $\mathbf{H}^{3}\in\mathbb{R}^{K_{o}^{3}\times\Phi^{3}_{h}\times\Phi^{3}_{w}}$, with $K_{o}^{3}$ to be the output channels of the third CNN, is reshaped to $\mathbf{H}^{cnn}\in\mathbb{R}^{\Phi^{cnn}_{h}\times\Phi^{cnn}_{w}}$, where $\Phi^{cnn}_{h} = \Phi^{3}_{h}$ and $\Phi^{cnn}_{w} = K_{o}^{3}\cdot\Phi^{3}_{w}$. $\mathbf{H}^{cnn}$ is given as an input to the GRU of the $f_{rnn}$. 

\subsection{Gated recurrent unit for long temporal context identification}
The output features $\mathbf{H}^{cnn}$ of $f_{cnn}$ are likely to include multi-scale contextual information, encoding long temporal patterns and inter- and intra-class activity~\cite{drossos:2019:dcase}. To exploit this information, the baseline approach utilizes $f_{rnn}$, which is a GRU that gets as an input the $\mathbf{H}^{cnn}$. The input and output dimensionality of $f_{rnn}$ the same and equal to $\Phi^{cnn}_{w}$. 

In particular, the GRU of $f_{rnn}$ takes as an input the output of the last CNN block of the baseline approach $\mathbf{H}^{cnn}$ and processes each row $\phi^{cnn}_{h}$ according to the equations mentioned in the original paper~\cite{cho:2014:gru}. The output of $f_{rnn}$, $\mathbf{H}^{rnn}\in[-1, 1]^{\Phi^{cnn}_{h}\times\Phi^{cnn}_{w}}$ is given as an input to the classifier $f_{cls}$. 

\subsection{Classifier, loss, and optimization}
The classifier $f_{cls}$ gets as an input the output of $f_{rnn}$, $\mathbf{H}^{rnn}$. $f_{cls}$ consists of a learnable affine transform with shared weights through time, followed by a sigmoid non-linearity. The output of $f_{cls}$ is the output of the CRNN method, which is

\begin{equation}
    \hat{\mathbf{Y}} = f_{\text{cls}}(\mathbf{H}^{rnn})\text{.}
\end{equation}

$f_{\text{cnn}}$, $f_{\text{rnn}}$, and $f_{\text{cls}}$ are jointly optimized by minimizing the cross-entropy loss between $\hat{\mathbf{Y}}$ and $\mathbf{Y}$. 

\section{Proposed approach}\label{sec:proposed-approach}
In our method we replace the $f_{cnn}$ and $f_{rnn}$ with different types of convolutions. We replace the $f_{cnn}$ with depthwise separable convolutions, which result in smaller amount of parameters and increased performance~\cite{chollet:2017:cvpr,szegedy:2015:cvpr,szegedy:2016:cvpr,ioffe:2015:icml,howard:2017:mobilenets}. Additionally, we replace the $f_{rnn}$ with dilated convolutions, which have smaller amount of parameters, are based on CNNs, and can model long temporal context~\cite{holschneider:1990:springer,shensa:1992:tsp,yu:2016:MCA}. 

Specifically, our method also accepts as an input $\mathbf{X}\in\mathbb{R}^{T\times N}$ and the associated annotations for the activities of classes $\mathbf{Y}\in\{0,1\}^{T\times C}$. $\mathbf{X}$ is given as an input to a learnable feature extractor $f_{\text{dws}}$, consisting of cascaded 2D depthwise separable CNN (DWS-CNN) blocks. Each block has a 2D CNN based on depthwise separable convolution followed by a non-linearity, a normalization process, and a feature sub-sampling process. The output of $f_{\text{dws}}$ is given as an input to a temporal pattern identification module $f_{\text{dil}}$, which consists of a 2D CNN based on dilated convolution (DIL-CNN). $f_{\text{dil}}$ is followed by a classifier $f_{\text{cls}}$, which is the same classifier as in the baseline approach. The output of $f_{\text{cls}}$ for each of the $T$ feature vectors is the predicted activities for each of the $C$ classes $\hat{\mathbf{Y}}\in [0, 1]^{T\times C}$. Similarly to the baseline, during the inference process, the activities $\hat{\mathbf{Y}}$ are further binarized using a threshold of 0.5.

\subsection{Learnable feature extractor based on depthwise separable convolutions}\label{subsec:dws}
Based on~\cite{sifre:2014:phd} and for our $f_{dws}$, we employ the factorization of the spatial and cross-channel learning process described by Eq~\eqref{eq:conv}. We replace the 2D CNNs of the CRNN method with 2D DWS-CNNs, closely following the DWS-CNNs presented for the MobileNets model~\cite{howard:2017:mobilenets} and the hyper-parameters used in the CRNN architecture~\cite{cakir:2017:taslp}. Instead of using $\mathbf{\Phi}$ in a convolution with a single kernel $\mathbf{K}$ in order to learn spatial and cross-channel information, we apply, in series, two convolutions (i.e. the output of the first is the input to the second) using two different kernels. This factorization technique is termed as depthwise separable convolution, has been adopted to a variety of architectures for image processing (like the XCeption, GoogleLeNet, Inception, and MobileNets models), and has been proven to increase the performance while reducing the amount of parameters~\cite{chollet:2017:cvpr,szegedy:2015:cvpr,szegedy:2016:cvpr,ioffe:2015:icml,howard:2017:mobilenets}. 

Firstly, we apply $K_{\text{i}}$ kernels $\mathbf{K}^{\text{s}}\in\mathbb{R}^{K_{\text{h}}\times K_{\text{w}}}$ to each $\mathbf{\Phi}_{k_{\text{i}}}$ in order learn the spatial relationships of features in $\mathbf{X}$ as

\begin{align}\label{eq:depthwise-conv}
    \mathbf{D}_{k_{\text{i}},t-K_{\text{h}}, n-K_{\text{w}}} =& (\mathbf{K}^{\text{s}}_{k_{\text{i}}} * \mathbf{X}_{k_{\text{i}}})({t-K_{\text{h}}, n-K_{\text{w}}})\nonumber\\
    =&\sum\limits_{k_{\text{h}}}^{K_{\text{h}}}\sum\limits_{k_{\text{w}}}^{K_{\text{w}}}\mathbf{X}_{k_{\text{i}}, t-k_{\text{h}}, n-k_{\text{w}}}\mathbf{K}^{\text{s}}_{k_{\text{i}}, k_{\text{h}}, k_{\text{w}}}\text{,}
\end{align}

\noindent
where $\mathbf{D}_{k_{\text{i}}}\in\mathbb{R}^{\Phi_{h}'\times\Phi_{w}'}$. Then, we utilize $K_{\text{o}}$ kernels $\mathbf{k}^{\text{z}}_{k_{o}}\in\mathbb{R}^{K_{\text{i}}}$, with $\mathbf{K}=\{\mathbf{k}_{1}^{\text{z}},\mathbf{k}_{2}^{\text{z}},\ldots,\mathbf{k}_{K_{o}}^{\text{z}}\}$, and we apply them $\mathbf{D} = \{\mathbf{D}_{1}, \ldots,\mathbf{D}_{K_{\text{i}}}\}$, in order learn the cross-channel relationships, as

\begin{equation}\label{eq:pointwise-conv}
    \mathbf{H}_{k_{\text{o}},\phi_{h}', \phi_{w}'} = \sum\limits_{k_{\text{i}}}^{K_{\text{i}}}\mathbf{D}_{k_{\text{i}},\phi_{h}', \phi_{w}'}\mathbf{K}^{\text{z}}_{k_{\text{o}}, k_{i}}\text{.}
\end{equation}

\begin{figure}[!t]
    \centering
    \subfloat[The first step of depthwise separable convolution. Learning spatial information, using $K_{\text{i}}$ different kernels $\mathbf{K}^{s}$, applied to each $\mathbf{X}_{i}$.\label{subfig:depthwise-conv-a}]{%
       \includegraphics[width=.9\columnwidth]{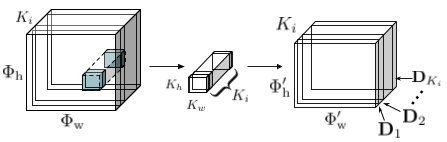}
     }
     \hfill
     \subfloat[The second step of depthwise separable convolution. Learning cross-channel information using $K_{\text{o}}$ different kernels $\mathbf{K}^{z}$. \label{subfig:depthwise-conv-b}]{%
       \includegraphics[width=0.45\textwidth]{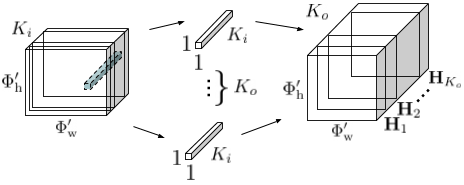}
     }
     \caption{The process of depthwise separable convolution}
     \label{fig:depthwise-conv}
\end{figure}

The resulting computational complexity and amount of parameters (omitting bias), for both processes in Eq.~\eqref{eq:depthwise-conv}~and~\eqref{eq:pointwise-conv}, are $O(K_{\text{h}}\cdot K_{\text{w}}\cdot K_{\text{i}}\cdot \Phi_{h} \cdot \Phi_{w} + K_{\text{i}}\cdot K_{\text{o}} \cdot \Phi_{h}' \cdot \Phi_{w}')$ and $K_{i}\cdot K_{h}\cdot K_{w} + K_{i}\cdot K_{o}$, respectively. Thus, the computational complexity~\cite{howard:2017:mobilenets} and amount of parameters are both reduced by $K_{\text{o}}^{-1} + (K_{\text{h}}\cdot K_{\text{w}})^{-1}$ times. The process of depthwise convolution is illustrated in Figure~\ref{fig:depthwise-conv}, with the first step in Figure~\ref{subfig:depthwise-conv-a} and the second in Figure~\ref{subfig:depthwise-conv-b}.

According to the baseline approach, we use three blocks of DWS-CNNs, where each block consists of a DWS-CNN, followed by a rectified linear unit (ReLU), a batch normalization process, and a max pooling operation across the dimension of features $\Phi_{w}$. $\mathbf{H}^{3}$ is the output of the third DWS-CNN block, which is given as an input to $f_{\text{dil}}$.

\subsection{Dilated convolutions}
Contrary to the baseline approach, we employ $f_{dil}$ in order to exploit the long temporal patterns in $\mathbf{H}^{3}$. $f_{dil}$ is based on 2D dilated convolutions, which are capable to aggregate and learn multi-scale information and have been used previously in image processing tasks~\cite{holschneider:1990:springer,shensa:1992:tsp,yu:2016:MCA}. 

A dilated 2D CNN (DIL-CNN) consists of $K'_{o}$ kernels $\mathbf{K}'\in\mathbb{R}^{K'_{\text{i}} \times K'_{\text{h}} \times K'_{\text{w}}}$ and bias vectors $\mathbf{b}'\in\mathbb{R}^{K'_{o}}$. Similarly to the typical CNN described in Section~\ref{subsec:dws}, $K'_{i}$ and $K'_{o}$ are the input and output channels of the DIL-CNN, and $K'_{h}$ and $K'_{w}$ are the height and width of the kernel for each channel. Each $\mathbf{K}'$ is applied to the input of DIL-CNN $\mathbf{\Psi}\in\mathbb{R}^{K'_{i}\times\Psi_{h}\times\Psi_{w}}$ to obtain the output $\mathbf{H}'\in\mathbb{R}^{K_{o}'\times\Psi_{h}'\times\Psi_{w}'}$ of the DIL-CNN as

\begin{align}\label{eq:conv_dilated}
    \mathbf{H}'_{k'_{\text{o}}, \psi_{h}'-k'_{\text{h}}, \psi_{w}'-k'_{\text{w}}}&=(\mathbf{K}_{k_{\text{o}}'}'*\mathbf{\Psi})(k_{\text{i}}', \psi_{h}-\xi_{\text{h}} \cdot k_{\text{h}}', \psi_{w}-\xi_{\text{w}}\cdot k_{\text{w}}')\nonumber\\
    =&\sum\limits_{k_{\text{i}}'}^{K_{\text{i}}'}\sum\limits_{k_{\text{h}}'}^{K_{\text{h}}'}\sum\limits_{k_{\text{w}}'}^{K_{\text{w}}'}\mathbf{\Psi}_{k_{\text{i}'}, \psi_{\text{h}}-\xi_{\text{h}}\cdot k_{\text{h}}, \psi_{\text{w}}-\xi_{\text{w}} \cdot k_{\text{w}}'} \mathbf{K}_{k_{\text{o}}', k_{\text{h}}', k_{\text{w}}'}'\text{,}
\end{align}

\noindent
where $\xi_{h},\xi_{w}\in\mathbb{N}^{\star}$ are the dilation rates for the $K'_{h}$ and $K'_{w}$ dimensions of $\mathbf{K}'$. It should be denoted that for $\xi_{h} = \xi_{w} = 1$, Eq.~\eqref{eq:conv_dilated} boils down to Eq.~\eqref{eq:conv}, i.e. a typical convolution with no dilation.

\begin{figure}[!t]
    \centering
    \subfloat[Calculation of $\protect\mathbf{H}'_{k_{o}', \psi_{h}',\psi_{w}'}$]{%
     \includegraphics[width=.45\columnwidth, trim={1cm 1.35cm 0 0cm},clip]{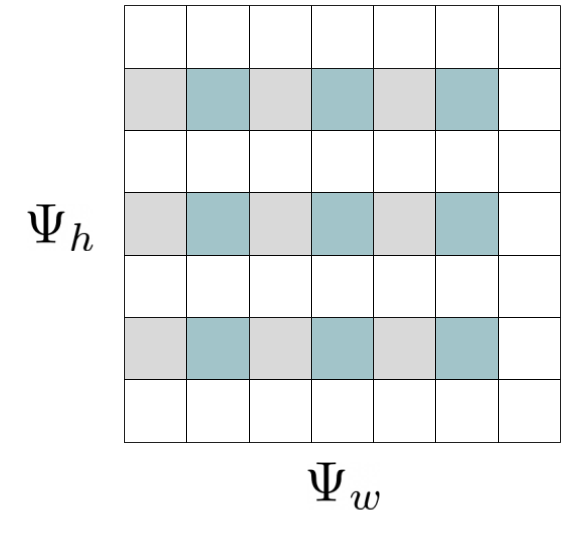}
     }~
    \hfill
     \subfloat[Calculation of $\protect\mathbf{H}'_{k_{o}, \psi_{h}',\psi_{w}'+1}$]{%
       \includegraphics[width=.45\columnwidth, trim={1cm 1.35cm 0 0cm},clip]{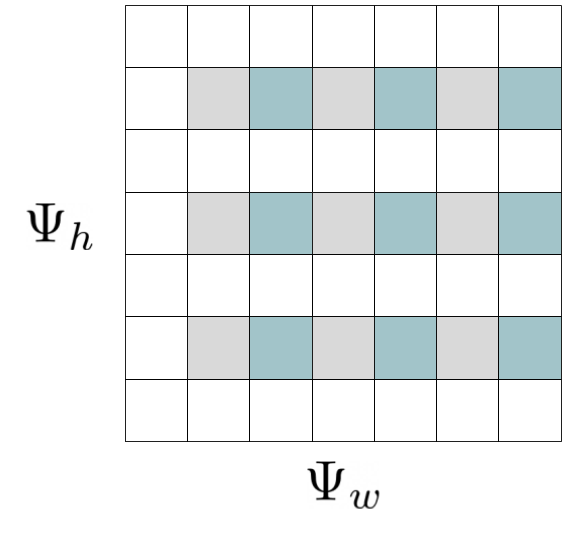}
     }
     \caption{Illustration of the process described in Eq.~\eqref{eq:conv_dilated} using $\xi_{h}=\xi_{w}=2$ and calculating two consecutive elements of $\mathbf{H}'_{k_{o}', \psi_{h}'}$. Squares coloured with cyan signify the elements participating at the calculations for $\protect\mathbf{H}'_{k_{o}', \psi_{h}',\psi_{w}'}$, and coloured with grey are the elements for $\protect\mathbf{H}'_{k_{o}', \psi_{h}',\psi_{w}'-1}$.}
     \label{fig:dilated-conv}
\end{figure}

The dilation rates, $\xi_{h}$ and $\xi_{w}$, multiply the index that is used for accessing elements from $\mathbf{\Psi}$. This allows a scaled aggregation of contextual information at the output of the operation~\cite{yu:2016:MCA}. Practically, this means that the resulting features computed by using DIL-CNN (i.e. $\mathbf{H}'$) are calculated from a bigger area, resulting into modelling longer temporal context. The growth of the area that $\mathbf{H}'$ is calculated from, is equal $\xi_{h}\cdot\xi_{w}$. The process described by Eq.~\eqref{eq:conv_dilated} is illustrated in Figure~\ref{fig:dilated-conv}.

We use DIL-CNN to replace the recurrent neural networks that efficiently model long temporal context and inter- and intra-class activities for SED. Specifically, our $f_{dil}$ has $K'_{i}=K_{o}$, takes as an input the output of $f_{dws}$, $\mathbf{H}^{L}$, and outputs $\mathbf{H}'$, as

\begin{align}
    &\mathbf{H}' = f_{dil}(\mathbf{H}^{L})\text{, and}\label{eq:dil}\\
    &\mathbf{H}^{dil} = BNorm(ReLU(\mathbf{H}'))\text{.}
\end{align}

\noindent
Finally, $\mathbf{H}^{dil}$ is reshaped to $\Psi_{h}'\times(K_{o}\cdot\Psi_{w}')$ and given as an input to the classifier of our method, which is the $f_{cls}$ of the baseline approach.  

\section{Evaluation setup}\label{sec:evaluation-setup}
To assess the performance of each of the proposed replacements and their combination, we employ a freely available SED dataset and we compare the performance of the CRNN and each of our proposed replacements. The code for all the models and the evaluation process described in this paper, is freely available online\footnote{\url{https://github.com/dr-costas/dnd-sed}}.

\subsection{Baseline system and models}
We employ four different models, $\text{Model}_{\text{base}}$, $\text{Model}_{\text{dw}}$, $\text{Model}_{\text{dil}}$, and $\text{Model}_{\text{dnd}}$. $\text{Model}_{\text{base}}$ is our main baseline and consists of three CNN blocks, followed by a GRU, and a linear layer acting as a classifier. Each CNN block consists of a CNN with 256 channels, square kernel shape of $\{5, 5\}$, stride of $\{1, 1\}$, and padding of $\{2, 2\}$, followed by a ReLU, a batch normalization, a max pooling, and a dropout of 0.25 probability. The max pooling operations have kernels and stride of $\{1, 5\}$, $\{1, 4\}$, and $\{1, 2\}$. The GRU has 256 input and output features, and the classifier has 256 input and 16 output features. 

For our second model, Model\textsubscript{dws}, we replace the CNN blocks at CRNN with $f_{dws}$, so we can assess the benefit of using DWS-CNNs instead of typical 2D CNNs. To minimize the factors that will have an impact to possible differences between our proposed method and the employed baseline, for our $f_{dws}$ we adopted the same kernel shapes, strides, and padding for the $\mathbf{K}_{\text{s}}$ kernels, as in the Model\textsubscript{base}. That is, all $K_{o}$, $K_{h}$, and $K_{w}$ of $f_{dws}$ have the same values as the corresponding ones in $\text{Model}_{\text{base}}$. The same stands true for stride and padding, and all hyper-parameters of max-pooling operations. 

At the third model, $\text{Model}_{\text{dil}}$, we replace the GRU in $\text{Model}_{\text{base}}$ with the $f_{dil}$, so we can assess the benefit of using DIL-CNN instead of an RNN. Since there are no previous studies using DIL-CNNs as a replacement for RNNs and for SED, we opt to keep the same amount of channels at the DWS-CNNs but perform a grid search on $K'_{h}$, $K'_{w}$, and $\xi_{h}$. Specifically, we employ four different kernel shapes $(K'_{h}, K'_{w})\in\{(3, 3), (5, 5), (7, 7)\}$. We denote the different shapes of kernels with an exponent, e.g. $\text{Model}_{\text{dil}}^{3}$ for the model having an $f_{dil}$ with a kernel of shape of $\{3, 3\}$, or $\text{Model}_{\text{dnd}}^{7}$ for the model having $f_{dws}$ and an $f_{dil}$ of kernel with shape $\{7, 7\}$. Because we want to assess the effect of using a different time-resolution for capturing inter- and intra-event patterns with the DIL-CNN, we use $\xi_{w} = 1$ and $\xi_{h}\in\{1, 10, 50, 100\}$. That is, we apply dilation only on the time dimension and not on the dimension of features. Though, to keep the time dimension intact (i.e. to have $\Psi_{h}'=T$), we use zero padding at the time dimension. Specifically, we use a padding equal to $\xi_{h}$ for kernel shape of $(3, 3)$, a padding equal to $2\cdot\xi_{h}$ for $(5, 5)$ kernel, $3\cdot\xi_{h}$ for the $(7, 7)$ kernel, and $5\cdot\xi_{h}$ for the $(11, 11)$ kernel. We use no padding at the feature dimension for the $f_{dil}$. Must be noted that when $\xi_{h} = 1$ then $f_{dil}$ is a typical 2D CNN and, thus, we also assess the effect of replacing the RNN with a typical 2D CNN. We also denote the employed dilation in the exponent, e.g. $\text{Model}_{\text{dil}}^{3|50}$ or $\text{Model}_{\text{dnd}}^{7|1}$. 

Finally, the $\text{Model}_{\text{dnd}}$ is our complete proposed method, where we replace both the CNN blocks and the GRU from the $\text{Model}_{\text{base}}$, with the $f_{dws}$ and $f_{dil}$, respectively. For complete assessment of our proposed method, we follow the same grid search on on $K'_{h}$, $K'_{w}$, and $\xi_{h}$, as we perform for $\text{Model}_{\text{dil}}$. 

\begin{table*}[!ht]
 \caption{Quantitative results from evaluating the effect of using depth-wise separable ($\text{Model}_{\text{dwd}}$) or dilated ($\text{Model}_{\text{dil}}$), or both ($\text{Model}_{\text{dnd}}$) convolutions as modifications to the baseline CRNN architecture ($\text{Model}_{\text{base}}$). Average (mean) values of the F1 score ($\overline{F}_{1}$, higher the better) and the error rate ($\overline{\text{ER}}$, lower the better) are reported over the ten repetitions. The number of parameters is denoted by $N_{P}$ and the average (and standard deviation, STD) time, in seconds, required for an epoch by $\overline{E}_{T}$ ($\pm$STD). N/A denotes a non applicable parameterization. Bold faced elements denote the best reported performance for classification and computational performance.} 
 \begin{center}
 \resizebox{\linewidth}{!}{%
 \begin{tabular}{cccc|cc|cc}
  \multicolumn{4}{c}{} &
  \multicolumn{2}{c|}{\textbf{SED Performance}} & \multicolumn{2}{c}{\textbf{Computational Performance (mean$\pm$STD)}} \\

  $\text{Model}_{*}$ & DWS & $\xi_{\text{h}}$ & ($K'_\text{h}, K'_\text{w},$) & $\overline{F}_{1}$ & $\overline{\text{ER}}$ & $N_{P}$ 
  & $\overline{E}_{T}$ 
  \\
  \hline

  \multirow{1}{*}{$\text{base}$} 
  & $\times$ & N/A & N/A 
  & 0.59
  & 0.54
  & 3.68M 
  & 49.4 ($\pm 11.8$) 
  \\
\hline
  \multirow{12}{*}{$\text{dil}$} 
  & $\times$ & 1 & ($3 \times 3$)  
  & 0.60
  & 0.54
  & 3.81M 
  & 14.1 ($\pm 0.06$) 
  \\
  
   & $\times$ & 10 & ($3 \times 3$) 
   & 0.61
   & 0.53
   & 3.81M
   & 14.1 ($\pm 0.11$) 
   \\
   
   & $\times$ & 50 & ($3 \times 3$)
   & 0.62
   & 0.51
   & 3.81M 
   & 14.1 ($\pm 0.07$) 
   \\
      
  & $\times$ & 100 & $3 \times 3$)
  & 0.61
  & 0.53
  & 3.81M
  & 14.5 ($\pm 0.08$)
  \\
   
  & $\times$ & 1 & ($5 \times 5$)
  & 0.60
  & 0.54
  & 3.81M
  & 20.7 ($\pm 0.09$)
  \\

  &$\times$ & 10 & ($5 \times 5$)
  & 0.63
  & 0.51
  & 3.81M
  & 18.2 ($\pm 0.25$)
  \\
   
  & $\times$ & 50 & ($5 \times 5$)
  & 0.60
  & 0.52
  & 3.81M
  & 18.5 ($\pm 0.07$)
  \\
   
  & $\times$ & 100 & ($5 \times 5$)
  & 0.58
  & 0.56
  & 3.81M
  & 18.5 ($\pm 0.08$) 
  \\
   
  & $\times$ & 1 & ($7 \times 7$)
  & 0.60
  & 0.54
  & 3.64M
  & 12.2 ($\pm 0.06$)
  \\
   
  & $\times$ & 10 & ($7 \times 7$)
  & 0.62
  & 0.52
  & 3.64M
  & 12.2 ($\pm 0.07$) 
  \\
   
  & $\times$ & 50 & ($7 \times 7$)
  & 0.61
  & 0.52
  & 3.64M
  & 12.4 ($\pm 0.07$)
  \\
   
  & $\times$ & 100 &($7 \times 7$)
  & 0.58
  & 0.57
  & 3.64M
  & 12.4 ($\pm 0.07$)
  \\
  
  \hline
  \multirow{1}{*}{$\text{dws}$} & $\checkmark$ & N/A & ($3 \times 3$)  
  & 0.62
  & 0.50
  & 0.62M
  & 46.9 ($\pm 4.81$)
  \\
  
  \hline
  \multirow{12}{*}{$\text{dnd}$} & $\checkmark$ & 1 & ($3 \times 3$)
  & 0.59
  & 0.54
  & 0.74M
  & 13.0 ($\pm 0.06$)
  \\
   
  & $\checkmark$ & 10 & ($3 \times 3$)
  & 0.62
  & 0.51
  & 0.74M
  & 13.0 ($\pm 0.06$)
  \\
   
  & $\checkmark$ & 50 & ($3 \times 3$)
  & 0.61
  & 0.53
  & 0.74M
  & 13.0 ($\pm 0.10$)
  \\
   
  & $\checkmark$ & 100 & ($3 \times 3$)
  & 0.60
  & 0.53
  & 0.74M
  & 13.4 ($\pm 0.08$)
  \\
   
  & $\checkmark$ & 1 & ($5 \times 5$)
  & 0.59
  & 0.55
  & 0.74M
  & 20.1 ($\pm 3.63$)
  \\
   
  & $\checkmark$ & 10 & ($5 \times 5$)
  & 0.62
  & 0.52
  & 0.74M
  & 17.0 ($\pm 0.24$)
  \\
   
  & $\checkmark$ & 50 & ($5 \times 5$)
  & 0.62
  & 0.52
  & 0.74M
  & 17.4 ($\pm 0.01$)
  \\
   
  & $\checkmark$ & 100 & ($5 \times 5$)
  & 0.58
  & 0.56
  & 0.74M
  & 17.4 ($\pm 0.01$)
  \\
   
  & $\checkmark$ & 1 & ($7 \times 7$)
  & 0.60
  & 0.54
  & 0.58M
  & 11.4 ($\pm 4.45$)
  \\
   
  & $\checkmark$ & 10 & ($7 \times 7$) 
  & \textbf{0.63}
  & \textbf{0.50}
  & \textbf{0.58M}
  & \textbf{11.1} ($\pm 0.06$)
  \\
   
  & $\checkmark$ & 50 & ($7 \times 7$)
  & 0.61
  & 0.53
  & 0.58M
  & 11.2 ($\pm 0.17$)
  \\

  & $\checkmark$ & 100 & ($7 \times 7$)
  & 0.58
  & 0.57
  & 0.58M
  & 11.3 ($\pm 0.11$)
  \\
 \end{tabular}
 }
\end{center}\label{tab:resu}
\end{table*}

\subsection{Dataset and metrics}
We use the TUT-SED Synthetic 2016 dataset, which is freely available online\footnote{\url{http://www.cs.tut.fi/sgn/arg/taslp2017-crnn-sed/tut-sed-synthetic-2016}} and has been employed in multiple previous work on SED~\cite{cakir:2017:taslp,drossos:2019:dcase,huang:2018:iwaenc}. TUT-SED Synthetic consists of 100 mixtures of around eight minutes length with isolated sound events from 16 classes, namely alarms \& sirens, baby crying, bird singing, bus, cat meowing, crowd applause, crowd cheering, dog barking, footsteps, glass smash, gun shot, horse walk, mixer, motorcycle, rain, and thunder. The mixtures are split to training, validation, and testing split by 60\%, 20\%, and 20\%, respectively. The maximum polyphony of the dataset is 5. From each mixture we extract multiple sequences of $T=1024$ vectors, having $N=40$ log-mel band energies and using a hamming window of $\approx 0.02$ sec, with 50\% overlap. As the evaluation metrics we use F\textsubscript{1} score and error rate (ER), similarly to the original paper of CRNN and previous work on SED~\cite{cakir:2017:taslp,drossos:2019:dcase,huang:2018:iwaenc}. Both of the metrics are calculate on a per-frame basis (i.e. for every $t=1,2,\ldots,T$). Additionally, we keep a record of the training time per epoch for each model and for all repetitions of the optimization process, by measuring the elapsed time between the start and the end of each epoch. 

\subsection{Training and testing procedures}
We optimize the parameters of all models (under all sets of hyper-parameters) using the training split of the employed dataset, the Adam optimizer with values for hyper-parameters as proposed in the original paper~\cite{kingma:2015:adam}, a batch size of 16, and cross-entropy loss. After one full iteration over the training split (i.e. one epoch), we employ the same loss and measure its value on the validation split. We stop the optimization process if the loss on the validation split does not improve for 30 consecutive epochs and we keep the values of the parameters of the model from the epoch yielding the lowest validation loss. Finally, we assess the performance of each model using the testing split and the employed metrics (i.e. F\textsubscript{1} and ER). 

In order to have an objective assessment of the impact of our proposed method, we repeat 10 times the optimization for every model, following the above described process. Then, we calculate the average and standard deviation of the above mentioned metrics, i.e., F\textsubscript{1} score and error rate (ER). In addition to this, we report the number of parameters ($N_{P}$) and the necessary mean training time per epoch ($\overline{E}_{T}$), i.e., a full iteration throughout the whole training split. All presented experiments performed on an NVIDIA Pascal V100 GPU. 

\section{Results and discussion}\label{sec:results}
In Table~\ref{tab:resu} are the results from all conducted experiments, organized in two groups. The first one is termed as SED performance and regards the performance of each model and set of hyper-parameters for the SED task (i.e. F\textsubscript{1} and ER). The second group, termed as computational performance, considers the number of parameters and average time necessary for training ($N_{P}$ and $\overline{E}_{T}$), for each model and each set of hyper-parameters. The STD of F\textsubscript{1} and ER is in the range of $0$ to $0.02$ and omitted for clarity.

The baseline CRNN system, i.e. Model\textsubscript{base}, seems to perform better in classification only from $\text{Model}_{\text{dnd}}^{7|100}$. In every other case, Model\textsubscript{base} yields worse classification performance. This indicates that our proposed changes can, in general, result to better classification performance when compared to the baseline system. Regarding the computational performance, can be seen that there are specific sets of hyper-parameters that result to models with more parameters from  Model\textsubscript{base}. Specifically, $\text{Model}_{\text{dil}}^{3}$ and $\text{Model}_{\text{dil}}^{5}$ with all $\xi_{h}$, have more parameters than Model\textsubscript{base}. This increase in $N_{P}$, though, is not attributed on the difference of the amount of parameters between $f_{dil}$ of Model\textsubscript{dil} and the GRU of Model\textsubscript{base}, but on the amount of parameters that the classifier has. In the case of Model\textsubscript{base}, the output of the GRU had dimensions of $1024\times256$. The classifier has shared weights through time, thus the amount of its input features is 256. But, in the case of $\text{Model}_{\text{dil}}^{3}$ and $\text{Model}_{\text{dil}}^{5}$, the dimensionality of the input to the classifier, i.e. $\mathbf{H}^{dil}$, is $256\times1024\times\Psi_{w}'$, where $\Psi_{w}'$ is inverse proportional to the size of the kernel of $f_{dil}$. After reshaping $\mathbf{H}^{dil}$ to $1024\times(256\cdot\Psi_{w}')$, the amount of input features to the classifier is $K_{o}\cdot\Psi_{w}'$, which is considerably bigger than the $\text{Model}_{\text{base}}$ case, i.e. $1024\times256$. Finally, Model\textsubscript{base} needs more time (on average) per epoch compared to any other model and set of hyper-parameters in Table~\ref{tab:resu}. This clearly indicates that all of the proposed changes have a positive impact on the needed time per epoch, even in the case where $N_{P}$ is bigger. 

Comparing the impact of each of the changes (i.e. Model\textsubscript{dws} versus Model\textsubscript{dil}), we can see that adopting DWS-CNN can significantly increase the SED performance, yielding better F\textsubscript{1} and ER compared to Model\textsubscript{base} and Model\textsubscript{dil} (except $\text{Model}_{\text{dil}}^{5|10}$). Additionally, Model\textsubscript{dws} yields the lowest ER in total, but not the highest F\textsubscript{1}. Furthermore, Model\textsubscript{dws} has $N_{P}=0.62$ M, significantly less than any Model\textsubscript{dil} and the Model\textsubscript{base}. The decrease in the amount of parameters and the increase in the performance when using the $f_{dws}$ is in accordance with previous studies that adopted DWS-CNN~\cite{chollet:2017:cvpr,szegedy:2015:cvpr,szegedy:2016:cvpr,ioffe:2015:icml,howard:2017:mobilenets}. Focusing on the Model\textsubscript{dil}, can be observed that the usage of dilation increases the classification performance. Specifically, in all kernel shapes, the $\xi_{h}=1$ (i.e. no dilation) yields the lowest F\textsubscript{1} and highest ER. Also, it is apparent that for $\xi_{h}\geq50$ the classification performance decreases. 

Finally, when both $f_{dws}$ and $f_{dil}$ are combined (i.e. Model\textsubscript{dnd}) it seems that there is a drop in the performance (compared to Model\textsubscript{dws}) for the (3, 3) and (5, 5) kernel shapes and for all $\xi_{h}$. But, for the case of $\text{Model}_{\text{dnd}}^{7|10}$, there is the highest F\textsubscript{1} score and by 0.02 second ER. Additionally, the specific $\text{Model}_{\text{dnd}}^{7|10}$ model needs the less average time per epoch and belongs to the group of models with the less parameters.

\section{Conclusion and future work}\label{sec:conclusions}
In this paper, we proposed the adoption of depthwise separable and dilated convolutions based 2D CNNs, as a replacement of usual 2D CNNs and RNN layers in typical SED methods. To evaluate our proposed changes, we conducted a series of experiments, assessing each replacement in separate and also their combination. We used a widely adopted method and a freely available SED dataset. Our results showed that when both DWS-CNN and DIL-CNN are used, instead of usual CNNs and RNNs, respectively, the resulting method has considerably better classification performance, the amount of parameters decreases by 80\%, and the average needed time (for training) per epoch decreases by 72

Although we conducted a grid search of the hyper-parameters, the proposed method is likely not fine tuned for the task of SED. Further study is needed in order to fine tune the hyper-parameters and yield the maximum classification performance for the task of SED. 

\section*{Acknowledgement}
Part of the computations leading to these results was performed on a TITAN-X GPU donated by NVIDIA. Part of the research leading to these results has received funding from the European Research Council under the European Union’s H2020 Framework Programme through ERC Grant Agreement 637422 EVERYSOUND. Stylianos I. Mimilakis is supported in part by the German Research Foundation (AB 675/2-1, MU 2686/11-1). The authors wish to acknowledge CSC-IT Center for Science, Finland, for computational resources.

\bibliographystyle{IEEEtran}
\bibliography{refs}

\end{document}